\documentclass{jpsj-suppl}

\usepackage{txfonts} %Please comment out this line unless the txfonts package is availabe in your LaTeX system.

\newcommand{\plotwid}{0.9}
\newcommand{\lepton}{\ell}
\newcommand{\proton}{\textrm{p}}
\newcommand{\xx}{\textrm{X}}
\newcommand{\yy}{\textrm{Y}}
\newcommand{\dt}{\textrm{TT}}
\newcommand{\ztt}{\vec{z}_\dt}
\newcommand{\dptt}{\delta p_\dt}
\newcommand{\enu}{E_\textrm{rec}^\nu}
\newcommand{\sob}{\textrm{S}/\textrm{B}}

\title{Reconstruction of Energy Spectra of Neutrino Beams Independent of Nuclear Effects: Prospects for Current Experiments}

\author{Xianguo \textsc{Lu}$^{1}$}

\inst{$^{1}$ Department of Physics, Oxford University, Oxford, Oxfordshire, United Kingdom}

\email{Xianguo.Lu@physics.ox.ac.uk}

\recdate{\today}

\abst{The energy spectrum of a neutrino beam in the few-GeV region is free of uncertainties from nuclear effects when reconstructed via neutrino-hydrogen interactions. On  a multinuclear  (hydrogen containing) target such interactions  can be extracted using  transverse kinematic imbalance. We discuss the prospects of this technique for current experiments.}

\kword{Neutrino-hydrogen interaction, nuclear effects, energy reconstruction}

\begin{document}
\maketitle

\section{Introduction}

Hydrogen as target is advantageous for the  study of neutrino properties because it is not subject to  nuclear effects. However, its application has been tempered by technical difficulties. The last hydrogen bubble chamber for neutrino experiments    was BEBC at CERN before the mid-1980s~\cite{Allen:1985ti, Agashe:2014kda}. Due to safety concerns, in the last 30 years there has been no new measurement of neutrino interactions on pure hydrogen. 

Neutrino-nucleus (excluding hydrogen) interactions  do not provide a satisfactory alternative in the few-GeV region because  individual nucleons  are resolved but there is no experimental control of their kinematics. The effects associated with nuclear targets are highly convolved: uncertainties from the binding energy, Fermi motion, multinucleon correlations and final-state interactions (FSIs) are present in different interaction channels for all nuclear targets. These nuclear effects lead to an imprecise knowledge  of the neutrino energy spectrum, the latter in turn preventing  a direct measurement of the former. A possible  disentanglement between  the two  via  identifying nuclear effects with transverse kinematics imbalance  has been recently proposed~\cite{Lu:2015hea,  Lu:2015tcr}. Its application on  multinuclear (hydrogen containing) targets enables an extraction of neutrino-hydrogen interactions.

The technique of  double-transverse momentum imbalance~\cite{Lu:2015hea} in the neutrino charged-current (CC) resonance channel   can separate \emph{hydrogen events} in a multinuclear sample, providing a practical way of using pure hydrogen as target and therefore allowing a reconstruction of neutrino energy spectra independent of nuclear effects. The separation also enables a measurement of the neutrino-hydrogen cross section and a direct access to   nuclear effects. A detailed discussion can be found in Ref.~\cite{Lu:2015hea}. In the following, practical variations  of this technique   and  the prospects for current experiments are discussed.

\section{The double-transverse kinematic imbalance $\dptt$}

\begin{figure}%[!ht]
\begin{center}
\includegraphics[width=0.65\columnwidth]{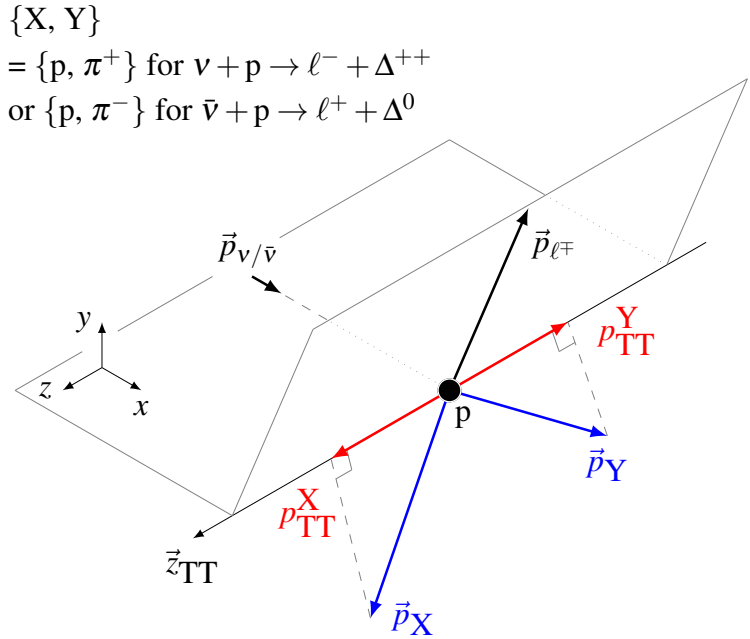}
\caption{Schematic illustration of the double-transverse kinematics.}\label{fig:dpttdef}
\end{center}
\end{figure}

Consider a lepton ($\lepton$)-proton ($\proton$) interaction producing three charged particles,
\begin{align}
\lepton+\proton\to\lepton^\prime+\xx+\yy,\label{eq:lxy}
\end{align}
where $\lepton^\prime$, $\xx$ and $\yy$ denote the final-state lepton and hadrons, respectively. The leading order realization in the Standard Model is the neutrino ($\nu/\bar{\nu}$) CC $\Delta$(1232)-resonance  production, 
\begin{align}
\nu+\proton&\to\lepton^-+\Delta^{++}\to\lepton^-+\proton+\pi^+,\label{eq:chan}
\end{align}
and
\begin{align}
\bar{\nu}+\proton&\to\lepton^++\Delta^{0}\to\lepton^++\proton+\pi^-,\label{eq:antichan}
\end{align}
where $\pi^\pm$ are the charged pions. As is shown in Fig.~\ref{fig:dpttdef},  on defining the double-transverse axis $\ztt$ perpendicular to both initial- and final-state lepton momenta,
\begin{align}
\ztt\sim\vec{p}_{\nu/\bar{\nu}}\times\vec{p}_{\lepton^\mp}, \label{eq:zdef}
\end{align}
 the momentum of the resonance, $\vec{p}_\proton+\vec{p}_{\pi^\pm}$, is projected onto $\ztt$ to define the double-transverse momentum imbalance:
\begin{align}
\dptt&\equiv\left(\vec{p}_\proton+\vec{p}_{\pi^\pm}\right)\cdot\ztt \\
&\equiv p^\proton_\dt + p^{\pi^\pm}_\dt,
\end{align}
where $p^{\proton,\,\pi^\pm}_\dt$ are implicitly defined. $\dptt$ equals 0 for a hydrogen target, and spreads over several hundred MeV for a nuclear target. This distinct feature enables hydrogen events to be extracted from  multinuclear targets. 

Depending on the physics objective, variations of the definitions above allow certain flexibility in practice. 
\begin{enumerate}
\item To extract hydrogen events:
\begin{enumerate}
\item The exclusivity of the final state rather than the intermediate particle production is relevant. Therefore, taking into account  realizations of Eq.~\ref{eq:lxy} at subleading orders, the interactions (Eqs.~\ref{eq:chan} and~\ref{eq:antichan}) can be  generalized to  include all channels  with an exclusive $\lepton^\mp\proton\pi^\pm$ final state. 
\item\label{it:zttdef} The final-state particle used to define $\ztt$ can be chosen arbitrarily. For a given detector, because the momentum reconstruction quality depends on the particle type, optimization of the $\dptt$ resolution is therefore possible via varying the particle type used for $\ztt$. 
\end{enumerate}
\item To obtain the neutrino energy spectrum:
\begin{enumerate}
\item\label{it:longE} Once the hydrogen events are extracted, the neutrino energy can be calculated   as the sum  of the final-state energy or longitudinal momenta, the choice between which depends on the calorimetry, tracking and particle identification (PID) performance of the detector. 
\item If the detector resolution does not allow an  event-by-event selection of the hydrogen events, the following method provides an  alternative: Calculate the neutrino energy following~\ref{it:longE} above for each event in the multinuclear sample, restrict  the sample in a region of the reconstructed neutrino energy $\enu$, i.e. bin in $\enu$, and then  in each bin extract the hydrogen yield by statistical background subtraction from the multinuclear $\dptt$ distribution. Across all bins, the cross section-normalized yield   is the spectrum independent of nuclear effects.  Major uncertainties come from the accuracy of the background subtraction and the detector resolution of $\enu$ from the hydrogen events. (Nuclear events  cause secondary bias in the mean $\enu$ of each bin.) This method can be generalized by changing the binning variable, for example to the virtuality $Q^2$,  to obtain the corresponding differential hydrogen cross section.

\end{enumerate}
\item To measure nuclear effects:
\begin{enumerate}
\item Nuclear effects can be studied from a multinuclear sample, or directly from a pure nuclear target. The former provides a direct access to pure nuclear effects by canceling out  contributions at the nucleon level; the latter is useful for determining the nuclear background shape in the spectral measurement. 
\item The definition of $\ztt$ (Eq.~\ref{eq:zdef}) does not depend on the kinematics of the final-state hadrons, and therefore the distribution of $\dptt$ is symmetric about 0. Since the positive direction of $\ztt$ can be chosen arbitrarily, it can be fixed in such a way that  $p_\dt^\proton$ is always positive. The $\dptt$ distribution thereby is asymmetric and sensitive to the difference between proton and pion FSIs.
\end{enumerate}
\end{enumerate}
In addition, the advantage of Variations \ref{it:zttdef} and \ref{it:longE} is that, no PID is needed once the momenta are determined.

\section{Prospects for measurements}\label{sec:pros}

In Ref.~\cite{Lu:2015hea} the reconstruction performance of $\dptt$ in a Monte Carlo (MC) simulation of the T2K ND280 detector~\cite{Abe:2011ks} is shown as a function of the neutrino energy.  Because ND280 was not optimized for an exclusive measurement in the CC resonance channel, its performance is projected with variable configurations in the following in order to estimate the impact of different detector designs on extracting hydrogen events. The variation of the performance as opposed to the  nominal value   is the emphasis. 

\begin{figure}%[!ht]
\begin{center}
%\subfigure[]
{\includegraphics[width=\plotwid\columnwidth]{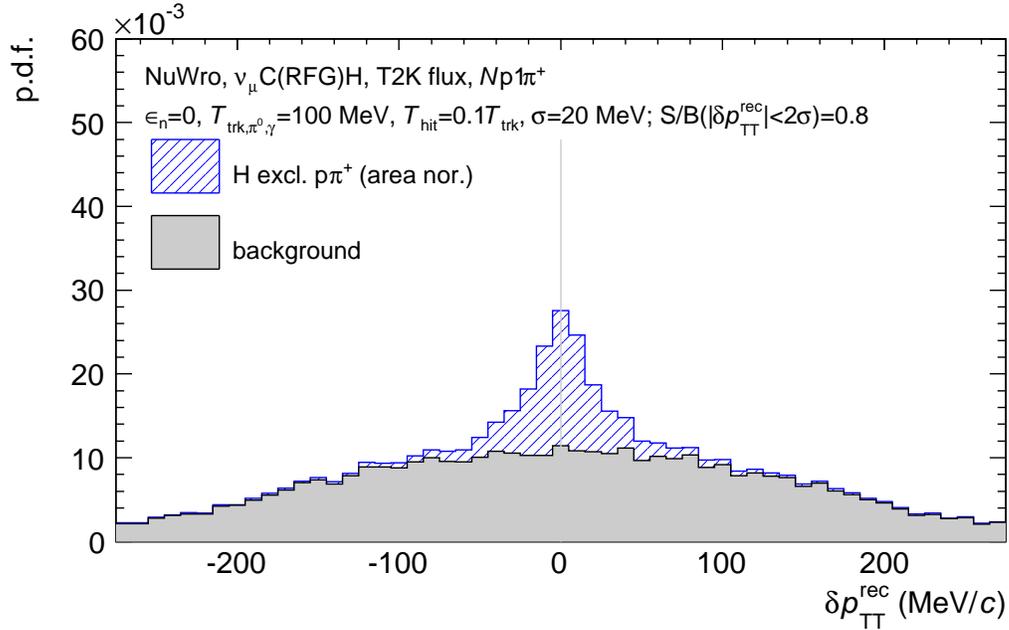}}
\caption{Distribution of the reconstructed $\dptt$ in the nominal configurations. The  hydrogen exclusive $\proton\pi^+$ signal (blue hatched) and the background (all other events) are stacked. The normalization is such that the area of the signal  equals 1. See text for explanations of the configurations.}\label{fig:t2k1}
\end{center}
\end{figure}

The performance projection is set up in a NuWro~\cite{Golan:2012wx}  simulation using the T2K flux~\cite{Abe:2011ks}. Muon neutrino ($\nu_\mu$) CC inclusive events (no multinucleon correlations) are generated on a CH target (nucleus state modeled as relativistic Fermi gas---RFG). ND280-like detector configurations (but with a $4\pi$ homogeneous acceptance) are used as the nominal set-up:
\begin{itemize}
\item CC muons are always tracked.
\item Neutrons are not detectable (i.e. neutron efficiency $\epsilon_\textrm{n}$ equals 0).
\item Neutral pions ($\pi^0$) and photons ($\gamma$) are registered if their kinetic energy is above  100 MeV (threshold $T_{\pi^0,~\gamma}=100$ MeV).
\item Charged particles are tracked if their kinetic energy is above 100 MeV (tracking threshold $T_\textrm{trk}=100$ MeV).
\item Untracked activities (\emph{hits}) are registered if a charged particle has kinetic energy above 10 MeV  (hit threshold $T_\textrm{hit}=10$ MeV) but below the tracking threshold.
\item The $\dptt$ resolution is 20 MeV/$c$ (Cauchy width $\sigma$)~\cite{Lu:2015hea}.
\end{itemize}
The nominal event selection  requires that at least 1 proton, exactly 1 $\pi^+$ and no other types of charged hadrons are tracked  (denoted by $N\proton 1\pi^+$). Transverse kinematics are then calculated with respect to the simulated neutrino direction. In practice, the neutrino  direction  can be reconstructed event by event as the direction from the mean decay point of the neutrino parents to the event vertex. This technique is most useful for  off-axis neutrinos to a near detector. Uncertainties in both neutrino direction reconstruction and tracking contribute to the resolution of the reconstructed transverse kinematics---in this study 20~MeV/$c$ for $\dptt$ is assumed. The assumed Cauchy shape is motivated by the characteristic  tails in the momentum smearing   caused by the multiple scattering  of a charged particle in detector material. The corresponding $\dptt$ distribution is shown in Fig.~\ref{fig:t2k1}. The signal to background ratio ($\sob$) in the $2\sigma$ window is 0.8. 

Due to FSIs, nuclear events can have nuclear emissions (products of nucleus excitation and break-up) in the final state.  Therefore $N\proton 1\pi^+$ events with more than 1 proton and/or any hits are  rejected to suppress the  nuclear background. Background from other channels, like deep inelastic scattering (DIS), is also partially removed by this cut. As the overall background is  reduced, the corresponding $\sob$ is increased to 0.9. Furthermore, since $\pi^0$ and $\gamma$ are produced in background channels,  an additional veto on them   increases the signal purity to $\sob=1.0$.

\begin{figure}%[!ht]
\begin{center}
%\subfigure[]
{\includegraphics[width=\plotwid\columnwidth]{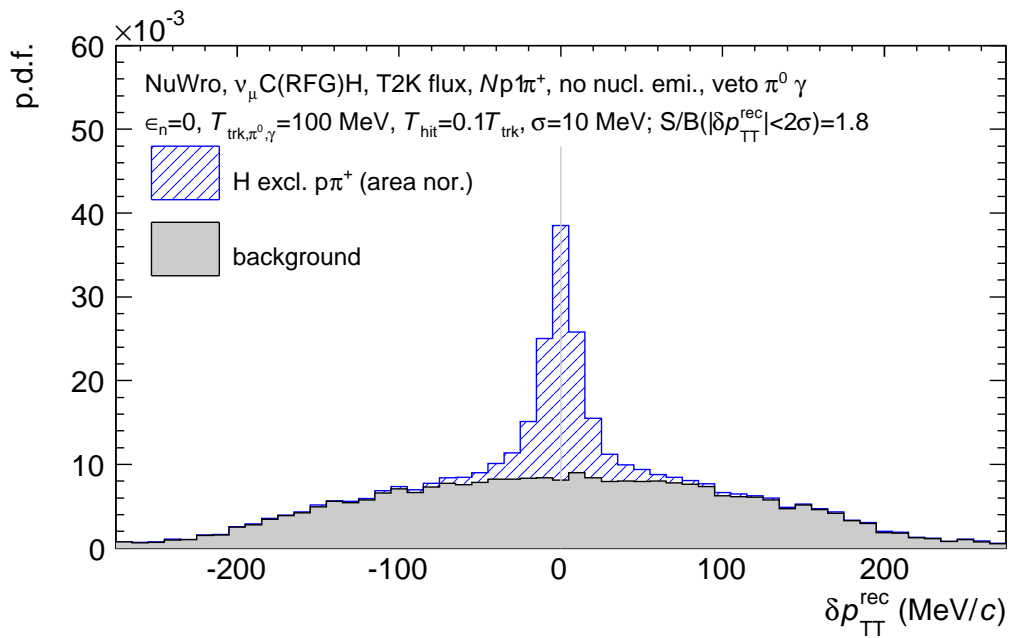}}
\caption{Reconstructed $\dptt$ distributions with an improved resolution. Nuclear emissions, $\pi^0$ and $\gamma$ are vetoed.}\label{fig:t2k4}
\end{center}
\end{figure}

Because the background, which is dominated by nuclear events, intrinsically distributes much more widely than the signal,  to first approximation the background shape under the signal peak is  flat, leading to $\sob\sim1/\sigma$.  In Fig.~\ref{fig:t2k4} where the resolution is improved by a factor of 2, i.e. $\sigma=10$~MeV, $\sob$ is increased to 1.8, demonstrating the  importance of the tracking resolution. For ND280, this can be  achieved, for example, by increasing the solenoid magnetic field from the current 0.2 T to 0.4 T (note that the ND280 magnet is capable to run at 0.8~T). 

Depending on the signal and background kinematics, the detection thresholds need to be optimized. A simple reduction of all thresholds by a factor of 2  does not have observable impact on the signal purity. 

\begin{figure}%[!ht]
\begin{center}
\includegraphics[width=\plotwid\columnwidth]{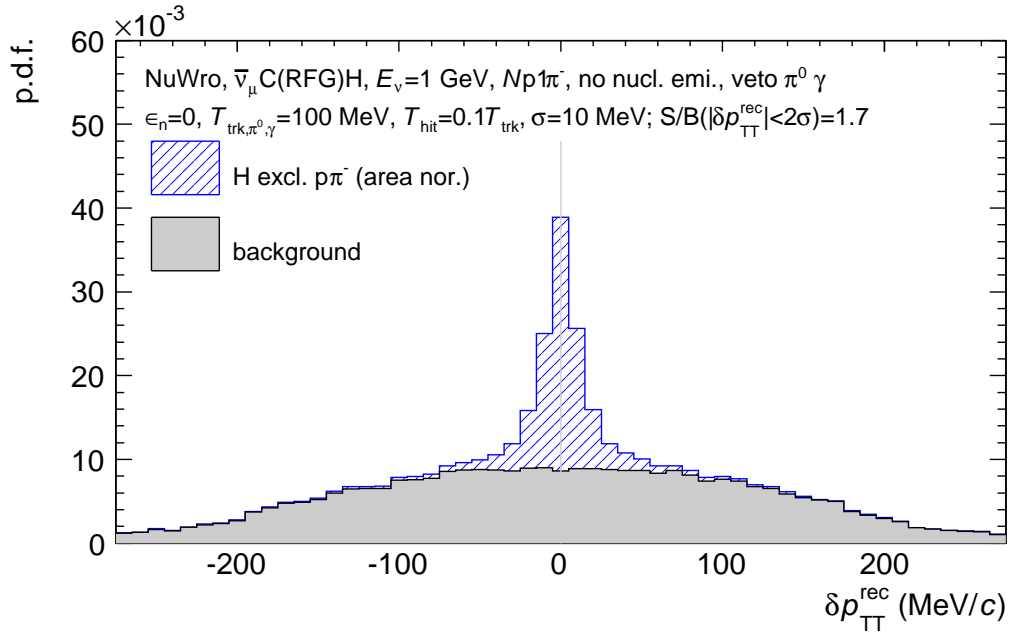}
\caption{Reconstructed $\dptt$ distributions for a $\bar{\nu}_\mu$ beam of energy of 1 GeV. The signal is the hydrogen exclusive $\proton\pi^-$ events.}\label{fig:antinu}
\end{center}
\end{figure}

Finally, the ND280 performance is projected for an anti-neutrino beam of energy of 1 GeV. A similar value of $\sob$ to the neutrino case is obtained (Fig.~\ref{fig:antinu}).

\section{Summary}

We reviewed the concept of the double-transverse momentum imbalance, a proposed technique to extract  hydrogen-neutrino interactions on multinuclear targets. Additional discussion on the variations for different applications is presented. A simple performance projection is set up based on ND280-like configurations, among which the tracking resolution is shown to be crucial. 

Hydrogen as target is attractive for the study of neutrino properties because of the lack of nuclear effects. With the proposed technique, nuclei in a   multinuclear target are useful because they provide a safe and convenient base for \emph{hydrogen doping}; uncertainties due to the nuclear effects can be eliminated by improving the tracking resolution---in fact, strong nuclear effects are preferred for an efficient removal of the background via, for example,  vetoing nuclear emissions.

In the current and  future liquid argon (LAr) TPC projects~\cite{Adams:2013qkq, Chen:2007ae, Anderson:2012vc, Adams:2013uaa, Antonello:2013ypa}, it would be very attractive to combine the superb tracking and calorimetry with the proposed use of hydrogen as target. Potential hydrogen doping in LAr TPCs would be desirable because the significant and yet much unknown nuclear effects of argon could be circumvented and even become useful. 

\section*{Acknowledgment}

I thank S.~Dolan and L.~Pickering for providing the numbers of the detection thresholds in Section~\ref{sec:pros}. The assistance on simulation provided by L.~Pickering is greatly acknowledged.

\end{document}